\documentclass[twocolumn,showpacs,preprintnumbers,amsmath,amssymb]{revtex4}
\usepackage{graphicx}% Include figure files
\usepackage{dcolumn}% Align table columns on decimal point
\usepackage{bm}% bold math

\begin{document}

%\preprint{APS/123-QED}

\title{Optical properties of boron-doped diamond}% Force line breaks with \\

\author{Dan Wu}
\author{Y. C. Ma}
\author{Z. L. Wang}
\author{Q. Luo}
\author{C. Z. Gu}
\author{N. L. Wang}\altaffiliation{Corresponding author}
\email{nlwang@aphy.iphy.ac.cn}

\affiliation{Beijing National Laboratory for Condensed Matter
Physics, Institute of
Physics, Chinese Academy of Sciences, Beijing 100080, China% with \\
}%

\author{C. Y. Li, X. Y. Lu, Z. S. Jin}
 %\homepage{http://www.Second.institution.edu/~Charlie.Author}
\affiliation{
State Key Laboratory for Superhard Materials, Jilin University, Changchun 130021, PR China% with \\
}%

\date{\today}% It is always \today, today,
             %  but any date may be explicitly specified

\begin{abstract}
We report optical reflectivity study on pure and boron-doped
diamond films grown by a hot-filament chemical vapor deposition
method. The study reveals the formation of an impurity band close
to the top of the valence band upon boron-doping. A schematic
picture for the evolution of the electronic structure with boron
doping was drawn based on the experimental observation. The study
also reveals that the boron doping induces local lattice
distortion, which brings an infrared-forbidden phonon mode at 1330
cm$^{-1}$ activated in doped sample. The antiresonance
characteristic of the mode in conductivity spectrum evidences the
very strong coupling between electrons and this phonon mode.
\end{abstract}

% PACS, the Physics and Astronomy
                             % Classification Scheme.
%\keywords{Suggested keywords}%Use showkeys class option if keyword
                              %display desired
\maketitle

Since diamond was used as an attractive material in many
applications for its high hardness and stable chemical property,
lots of studies have been done on diamond's transport and thermal
properties. Pure diamond is a wide-gap semiconductor with the band
gap of 5.5 eV.\cite{Walker,Saslow} As the inspiration of adding
impurity to semiconductor like silicon, the microwave plasma
chemical vapor deposition (MPCVD) method as well as high pressure
high temperature(HPHT) synthesizing skills are popularly used
recently to add dopants into pure diamond, with the purpose to
find out how impurities play roles in
diamond.\cite{Shiomi,Lagrange} The most attractive impurity now
should be boron as single electron acceptors. As the boron doping
level increasing, a gradual change from a semiconductor to a
metal, eventually to a superconductor occurs. Doped boron atoms
substitute for the carbon atoms when the concentration is low $\leq 0.5$ \verb+%+, and
occupy neutrally interstitial positions at the doping level $\sim$ 4 \verb+%+.
In low doping case, the substituted boron atoms are bonded to
neighboring carbon atoms in the $\emph{sp}^{3}$ environment, and
in the ground state the holes provided by boron atoms are bound in
one of the three fold degenerate impurity state with a binding
energy of 0.38 eV. At higher boron concentration, as the average
distance between boron atoms is close to the acceptor Bohr radius,
the metallic conduction appears, with the room temperature
conductivities of a few
$10^{2}~\Omega^{-1}$cm$^{-1}$.\cite{Werner,Bustarret,Nakamura}

The superconductivity of boron doped diamond at several Kelvins is
now the most interesting phenomenon.\cite{Ekimov,Takano} In order
to understand the origin of this phenomenon, a number of
first-principle calculations have been performed recently with
regard to the electronic structure, lattice dynamics, and the
electron-phonon coupling of the doped
system.\cite{Boeri,Lee,Blase,Xiang} It was suggested that the
superconductivity is mediated through the electron-phonon
interaction. Holes doped at the top of the zone-centered,
degenerate $\sigma$-bonding valence band couple strongly to the
optical bond-stretching modes, a mechanism similar to the one
causing the superconductivity in MgB$_2$.\cite{Boeri} However,
there is also a suggestion by Baskaran that an additional metallic
"mid-gap band" of a conducting 'self-doped' Mott insulator
contributes to the superconductivity.\cite{Baskaran}

In this paper, we study the optical property of lightly boron
doped diamond which has p-type semiconductor character and pure
diamond films in a wide range of frequency with the temperature
changing from 300 K to 10 K. A transfer of spectral weight from
interband transition to the low frequencies was clearly observed,
which provides experimental evidence for the evolution of the
electronic structure with boron doping. A schematic picture was
drawn based on the experimental observation. In addition, an
antiresonance phonon feature is clearly observed at low-frequency
$\sim$ 1330 cm$^{-1}$ for boron-doped compound, indicating (1) the
boron induced distortion is evident, which thus brings this
infrared-forbidden mode activated in doped sample; (2) strong
electron-phonon coupling.

The boron doped polycrystalline diamond thick film was grown using
an hot-filament chemical vapor deposition (HFCVD) method on
molybdenum substrates. The molybdenum substrates were
ultrasonically pretreated in an ethanol solution containing
diamond powder, followed by a -200 V bias voltage-assisted HFCVD
process. In addition to the CH$_{4}$ and H$_{2}$ used as reaction
gases, boron species were incorporated into the diamond films
during the growth process by bubbling the H$_{2}$ gas (10 sccm)
through the B(OCH$_{3}$)$_{3}$ liquid precursors at room
temperature. The total pressure was 50 Torr and the substrate
temperature was about 1073K measured by a thermocouple mounted on
the substrate. After deposition of 60 hours, the molybdenum
substrates were removed by a cooling technique, and a freestanding
diamond film with the thickness of about 100 $\mu$m was obtained.
The scanning electron microscopy image indicates that the average
diamond grain's size in the polycrystalline diamond thick film is
larger than 10 $\mu$m. X-ray diffraction (XRD) measurement
confirmed the pure phase and, to some degree, the preferred (111)
orientation of the diamond film as evidenced by the relatively
strong peak of the plane (111) in Fig. 1. The dc resistivity was
measured by standard four probe method. Fig. 2 displays the
resistivity vs T curves for pure and doped films. For the heavily
doped sample, the restivitiy value is much lower than the pure
sample, and becomes superconducting below $\sim$10 K.

\begin{figure}[t]
\includegraphics[width=7cm]{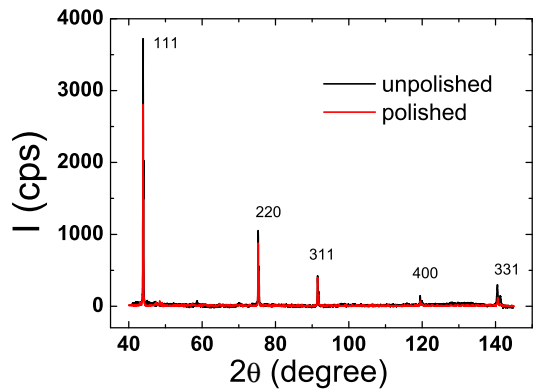}
\caption{\label{fig:resistivity}(color online)  The XRD patterns
with Cu K$_\alpha$ radiation ($\lambda$=0.154 nm) for a
boron-doped diamond film before and after polishing (black and red
lines, respectively).}
\end{figure}

The polycrystalline diamond films have very rough surfaces due to
the different growth orientations of diamond grains. For optical
reflectance measurement, the surface has to be polished. As the
diamond is a superhard material, polishing turns out to be very
difficult. We used a high rotate-speed (20000 round/min) diamond
grinding wheel to polish the sample and achieved a mirror-like
surface. Unfortunately, we found that polishing drastically
affects the properties of the sample. The superconducting sample
lost its superconductivity after being polished, meanwhile the
resistivity increases substantially. The $\rho$-T curve of the
same, but polished sample is also shown in Fig. 2. From the X-ray
diffraction patterns shown in Fig. 1, no discernible change in
peak positions could be found after polishing. However, the
testing of boron concentration by Hall effect with the polished
boron-doped sample shows that the boron concentration decreased to
1.3$\times$10$^{20}$cm$^{-3}$, which was much lower than the value
before polished, 7.32$\times$10$^{20}$cm$^{-3}$. This decreasing
of concentration crosses the critical value
n$_{c}$\cite{concentration} for the onset of superconductivity. It
is not clear why polishing so strongly affects the sample
properties. One possibility is that the boron concentration in the
sample is not homogeneous during the diamond film growth, the
outer surface region may have higher boron content, and the
polishing removes the outer surface of the sample which, as a
result, reduces the boron concentration. Further investigation on
this issue is needed. The near-normal incident reflectance spectra
were measured by a Bruker 66 v/s spectrometer in the range from 40
to 25000 cm$^{-1}$ and by a home-made grating spectrometer from
25000 cm$^{-1}$ to 50000 cm$^{-1}$. The sample was mounted on an
optically black cone in a cold-finger flow cryostat. An \emph{in
situ} overcoating technique was employed for reflectance
measurements.\cite{Homes} We performed Kramers-Kronig
transformation of R($\omega$) to obtain the optical conductivity
spectra. For the low frequency extrapolation, we use constant and
Hagen-Rubens relation for pure and doped samples, respectively. At
high frequency side, above the maximum frequency in the data file,
the reflectance is extrapolated as $\omega^{-1}$ up to 400,000
cm$^{-1}$ followed by a function of $\omega^{-4}$.

\begin{figure}[t]
\includegraphics[width=7cm]{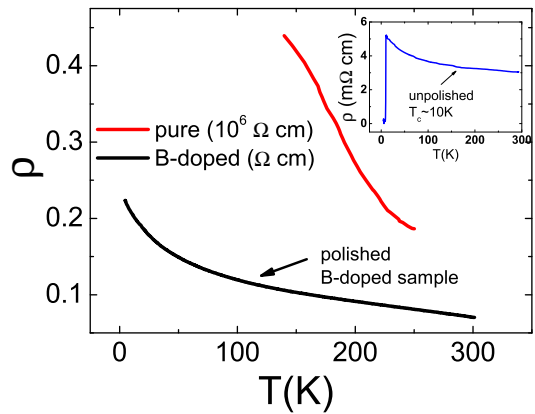}
\caption{\label{fig:resistivity}(color online) The dc resistivity
as a function of temperature for a pure and a boron-doped diamond
film. Unpolished boron doped sample exhibits a superconducting
transition near 10 K (Inset of the figure). The superconductivity
was lost unexpectedly after polishing.}
\end{figure}

Fig. 3 shows the optical reflectance spectra of pure and doped
samples measured at 10 K and 300 K. A weak temperature dependence
was found for the doped sample. The reflectance in the frequency
range of 150 - 2400 cm$^{-1}$ increases somewhat with decreasing
temperature, however, its reflectance below 150 cm$^{-1}$ or above
2400 cm$^{-1}$ decreases with decreasing temperature. Above 6400
cm$^{-1}$ the reflectance spectra are temperature independent. As
for the pure diamond sample, it shows featureless change over a
broad frequency range. The increase above 30000 cm$^{-1}$ is
attributed to the onset of interband transition. The low-$\omega$
reflectance decreases slightly with decreasing temperature, being
consistent with the insulating characteristic. As revealed clearly
in this figure, the major difference between the pure and B-doped
diamonds is a substantial increase of the low frequency
reflectance (below $\sim$5000 cm$^{-1}$) for the doped sample. The
low-$\omega$ spectral weight was transferred from high frequency
range above $\sim$5000 cm$^{-1}$ as well as in the interband
transition region. So, the reflectance data illustrate clearly
that B-doping leads to formation of electronic states at low
energies. Metallic conduction can be caused when those states are
enough to form a band crossing the Fermi level.

\begin{figure}[t]
\includegraphics[width=7cm]{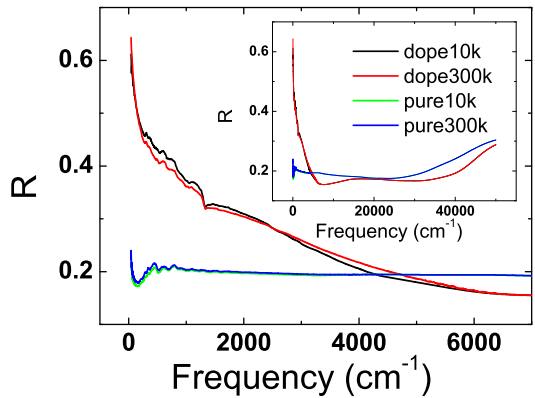}
\caption{\label{fig:refl}(color online) Frequency dependence of
reflectance spectra of pure and boron-doped diamond from 40
cm$^{-1}$ to 6400 cm$^{-1}$ at different temperature. The inset is
the whole spectra from 40 cm$^{-1}$ to 50000 cm$^{-1}$.}
\end{figure}

Fig. 4 shows the conductivity spectra in a broad frequency range.
For the pure diamond sample, the spectral weight at low frequency
is extremely low. A sharp increase in $\sigma(\omega$) appears at
frequency exceeding 30,000 cm$^{-1}$. This is due to the interband
transition from valence band to conduction band. For the B-doped
sample, on the other hand, an increase of spectral weight at
frequency below 7000 cm$^{-1}$ is evident. In the meantime, the
onset of interband transition shifts somewhat to higher
frequencies. Notably, the low energy excitations do not form a
Drude peak. The conductivity was severely suppressed at low
frequency, resulting in a broad peak centered at $\sim$3,000
cm$^{-1}$ (0.38 eV). This means that the states induced by
B-doping are still highly localized. We noticed that the peak
energy corresponds well to the boron acceptor levels locating at
0.38 eV from the top of valence band as determined from many other
experimental probes.\cite{Nakamura,Barnard} So, it just
corresponds to the interband transition from the top of the
valence band to the impurity states.

Based on the above experimental data, we can draw following
schematic picture for the evolution of the electronic states for B
doping diamond as shown in Fig. 5. The pure sample is a standard
semiconductor with the valence band completely filled and
conduction band completely empty. The Fermi level locates close to
the top of the valence band (Fig. 5(a)). B-doping creates the
acceptor impurity states (near the Fermi level) with a binding
energy of 0.38 eV.  When the doping levels are low, those energy
levels are isolated, and the impurity states are completely
localized (Fig. 5(b)). As the boron concentration increases, the
impurity energy levels broaden and form a band. At high enough
doping level, the impurity band may overlap with the top states of
valence band because of their very close energies (Fig. 5(c)).
Metallic conduction could be formed as long as the impurity band
crosses the Fermi level. Additionally, because the holes are
created in the valence band, a slight downward shift of Fermi
level likely occurs, which can explain the observed energy shift
of interband transitions. Obviously, for our polished B-doped
sample, it corresponds to the states between Fig. 5 (b) and (c).
Metallic response should be observed once the drastic effect
caused by polishing could be removed or reduced.

\begin{figure}[t]
\includegraphics[width=7cm]{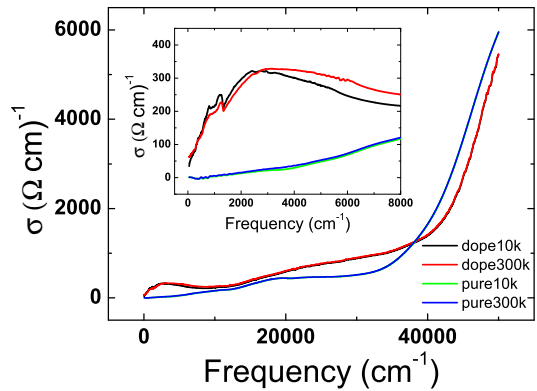}
\caption{\label{fig:sigma1} (color online)The optical conductivity
spectra of pure and boron-doped diamond over broad range of
frequencies. The inset is the expanded spectra from 40 cm$^{-1}$
to 8000 cm$^{-1}$.}
\end{figure}

The above picture shares some similarity with the high temperature
cuprate superconductors which are doped Mott insulators. For the
undoped parent compound, the lowest interband transition is the
charge transfer excitations, that is, the transition from occupied
O$_{2p}$ band to the empty upper Hubbard band of Cu 3d electrons.
When carriers, for example holes, are doped into the compound,
they starts to create some midgap states within the charge
transfer gap. In optical spectra, one can observe a transfer of
the spectral weight from the interband transition to the
mid-infrared band. With increasing doping, the mid-infrared band
shifts to lower frequency and finally merges with the Drude band
that appears when the concentration is higher than the critical
\emph{x}$_{c}$ for insulator-metal transition. Here, in the
boron-doped diamond system, the mid-gap states are just from those
boron acceptor impurity band. When the impurity band crosses the
Fermi level, the compound can becomes metallic. We believe that
the superconductivity originates from this impurity band. Baskaran
proposed that the impurity band has very strong electron
correlation, which could be split into lower and upper sub-bands,
in addition to some extra mid-gap states. If this is the case, one
would expect to see another interband transition from occupied
states to the upper Hubbard impurity band for boron doped sample.
The present experimental result does not support this scenario
because no such interband transition is observed.

Besides the change in electronic states with B-doping, there also
exists important difference between pure and doped sample with
regard to phonon spectra. We noticed that the doped sample has a
peculiar structure at about 1330 cm$^{-1}$ in reflectance and
conductivity spectra, while the pure sample does not show any
anomaly at this frequency. In the Raman spectra of pure diamond, a
very sharp zone-center phonon peak was observed at 1332
cm$^{-1}$.\cite{Ager} This Raman mode is obviously forbidden in
infrared as the reflectance is featureless in this particular
region. As the boron atoms substitute for carbon in crystal
structure, the distortion of vibrational eigenvectors brings
nonzero moments of electronic dipoles for most vibrational modes.
So the exceptional feature in doped one at $\sim$ 1330 cm$^{-1}$
should be explained as the activated infrared-forbidden mode by
impurity.

\begin{figure}[t]
\includegraphics[width=7cm]{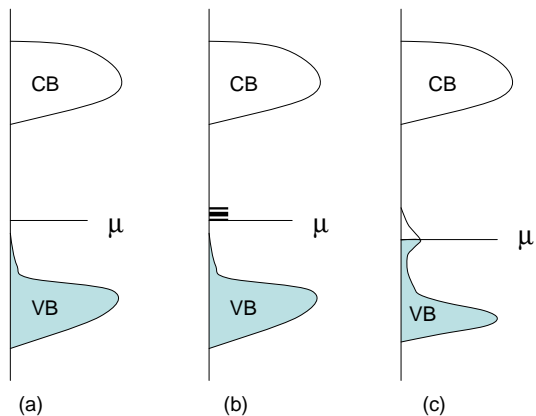}
\caption{\label{fig:bandpic}(color online) A schematic diagram for
the evolution of the electronic structure of diamond with
B-doping. (a) The pure diamond is a typical semiconductor. The
chemical potential is close to the top of the valence band. (b)
Boron-doping introduces acceptor impurity energy levels near the
top of the valence band. (c) With increasing boron doping, the
boron impurity levels form a band, and may overlap with the top of
the valence band. Metallic conduction occurs as long as the
impurity band crosses the chemical potential. }
\end{figure}

We noticed that this phonon mode has an anti-resonance
characteristic in the conductivity spectrum, namely, the feature
looks more like an asymmetric dip than a peak in the electronic
background of conductivity spectrum. Usually, a phonon displays a
peak structure centered at its characteristic frequency. The peak
can be asymmetric (Fano lineshape) due to an interaction of
lattice with the electrons. The rather obvious anti-resonance
feature is an indication of the strong electron-phonon coupling in
the compound. Such a phenomenon is more commonly seen in organic
conductor or superconductors as a result of significant
electron-molecular vibration coupling\cite{Jacobsen,Wang}. The
presence of such peculiar phonon structure unambiguously
illustrates the very strong electron-phonon coupling in present
doped diamond sample. In fact, the strong coupling of electrons
with such mode and its significant effect on the superconductivity
in B-doped diamond has been addressed in a number of theoretical
studies recently.\cite{Boeri,Lee,Blase,Xiang} Our infrared
response study provides experimental evidence for the strong
coupling of electrons with this mode.

In summary, we have performed optical reflectance measurements on
polished surfaces of both pure and boron-doped diamond films grown
by a hot-filament chemical vapor deposition method. Significant
difference has be found from the reflectance and conductivity
spectra in pure and boron-doped samples. The study revealed
clearly the formation of an acceptor impurity states near the top
of the valence band with boron doping. It is suggested that this
impurity band could overlap with the valence band at high enough
doping, and the superconductivity is originated from this impurity
band. The study also revealed the impurity's effect on phonon
vibration modes and a very strong electron-phonon coupling in the
doped compound.

We thank J. Y. Shen for her help in experiment. This work was
supported by the Ministry of Science and Technology of China (973
project No. 2006CB601002), the National High Technology
Development Program of China (Grant No. 2002AA325090), and the
National Science Foundation of China.

\end{document}